\documentclass[preprint, number]{elsarticle}
\usepackage{amsmath}
\usepackage{amsfonts}
\usepackage{graphicx}
\usepackage{dcolumn}
\usepackage{bm}
\usepackage{epstopdf}

\def\la{\langle}
\def\ra{\rangle}

\begin{document}

\title{Demographic Stochasticity versus Spatial Variation in the Competition between Fast and Slow Dispersers}

\author[umm]{Jack N. Waddell} 
\ead{seoc@umich.edu}
\author[ump,mctp]{Leonard M. Sander}
\ead{lsander@umich.edu}
\author[umm,ump,mctp,cscs]{Charles R. Doering} 
\ead{doering@umich.edu}

\address[umm] {Department of Mathematics, University of Michigan, Ann Arbor, Michigan 48109-1043}
\address[ump]{Department of Physics, University of Michigan, Ann Arbor, Michigan 48109-1040}
\address[mctp]{Michigan Center for Theoretical Physics,  University of Michigan, Ann Arbor, Michigan 48109-1040}
\address[cscs]{Center for the Study of Complex Systems,  University of Michigan, Ann Arbor, MI 48109-1107}
\cortext[cor1]{Corresponding author.}


\begin{abstract}
Dispersal is an important strategy that allows organisms to locate and exploit favorable habitats. The question arises: given competition in a spatially heterogeneous landscape, what is the optimal rate of dispersal? Continuous population models predict that a species with a lower dispersal rate always drives a competing species to extinction in the presence of spatial variation of resources. However, the introduction of intrinsic demographic stochasticity can reverse this conclusion. We present a simple model in which competition between the exploitation of resources and stochastic fluctuations leads to victory by either the faster or slower of two species depending on the environmental parameters. A simplified limiting case of the model, analyzed by closing the moment and correlation hierarchy, quantitatively predicts which species will win in the complete model under given parameters of spatial variation and average carrying capacity.
\end{abstract}


\maketitle

Dispersal plays an important role in population dynamics models. Dispersal may stabilize the species' populations \cite{Hastings-1982}, allow exploration of new habitats, and influence community structure \cite{Levine-2003}.  Despite the potential advantages of dispersal to a species, it may or may not be favored in a competitive environment.  That is, environmental factors influence whether faster dispersers or slower dispersers have a competitive advantage which in turn influences whether dispersal rates evolve toward faster or slower values. For example, spatial variability tends to reduce dispersal rates \cite{Hastings-1983, Dockery-1998, Hutson-2003} while temporal variability in the environment tends to increase dispersal rates \cite{Johnson-1990, Dieckmann-1999, Hutson-2001, Baskett-2007}.

Demographic stochasticity is known to be important in population dynamics.  In simple deterministic population models, the carrying capacity is often a stable fixed point, implying populations will remain for infinite time.  In a stochastic model, extinction, the only absorbing state, is certain at infinite time.  Even at finite times, demographic stochasticity leads to increased extinction risk  \cite{Lande-1993, Legendre-1999}, particularly when populations are small or fragmented.  The inclusion of stochasticity into non-linear mathematical models affects the mean dynamics, leading to, for example,  generation  \cite{Lande-1998} or exacerbation  \cite{Dennis-2002} of the Allee effect.  It can also actualize sustained cycles in a simple predator-prey model (i.e. Volterra) that are not present in the deterministic model \cite{McKane-2005}.

If demographic stochasticity can cause fundamental changes in the outcomes of simple population models, it is natural to wonder whether it can also cause significant changes in the competition of dispersers, and thus on the selection of dispersal rates.  In this study we develop a mathematical model in which spatial variation and demographic stochasticity are sufficient to select either a faster or a slower disperser in a competitive environment, and analytically estimate the boundary between the two outcomes. 


\section{Background}
The importance of demographic stochasticity is already evident in the classic study of Hamilton and May who examined a discrete population mean-field model where all sites were equivalent and stationary, but where there was an explicit cost of migration in the form of a fixed probability for migrating individuals to perish \cite{Hamilton-1977}.  In this scenario, Hamilton and May found a non-trivial evolutionarily stable rate of migration depending upon the probability of surviving migration, and later with Comins \cite{Comins-1980}, upon the carrying capacity per site as well.  Even in very poor environments, where the carrying capacity is unity and the probability of surviving migration is very small, the optimal migration rate goes to a non-zero constant, despite both spatial and temporal homogeneity of the environment.  This was interpreted as being due to a form of kin selection, specifically by reduction of kin competition.


Hastings \cite{Hastings-1983}, and later Dockery et al. \cite{Dockery-1998}, examined a deterministic, continuous population, continuous space model of $N$ competing species in a non-homogenous spatial environment modeled as
\begin{equation}
\frac{\partial u_i(x,t)}{\partial t} = D_i \nabla ^2 u_i(x,t) + u_i(x,t) \big(\gamma(x) - \sum_{j = 1}^N u_j(x,t)  \big)
\end{equation}
where $u_i(x,t)$ is the population of the $i$-th species, $D_i$ is the dispersal rate of the $i$-th species, and $\gamma(x)$ is the heterogeneous growth rate.  Under these conditions, they found that the species with the slower dispersal rate always drives a competing species to extinction.  Dockery et al conjectured that this generalizes beyond pair-wise competitions, implying that dispersal rates will always evolve towards zero in environments with any spatial variation.  

In essence, the result of Dockery et al states that a slowly dispersing species exploits favorable environments without wastefully exploring the landscape. The faster species does the complement, recklessly sending populations into poorer habitats.  This can be considered an \emph{implicit} cost of migration, rather than the \emph{explicit} cost of migration in Hamilton and May.  However, unlike Hamilton and May, the continuous population model disregards discreteness of and stochasticity in populations which drive migration rates upward.  Costs of migration, whether explicit or implicit, drive migration rates downward.

The Hamilton and May model (as analyzed by Comins et al) is consistent with the observation of Hastings and of Dockery et al in the continuum limit.  As the carrying capacity on each site approaches infinity, the evolutionarily stable dispersal rate approaches zero.  Although this was interpreted as the elimination of kin competition - if an infinite number of individuals can live on a site, then no one is competing for a position - it can be interpreted instead as an elimination of demographic stochasticity.  Indeed, the results of Hamilton and May can be derived by considering stochasticity without reference to kin selection \cite{Weissman-2009}.

Many others have also considered the effect of stochasticity on dispersal rate evolution.  Models studied might involve continuous \cite{Cadet-2003} or discrete (seasonal or generational) time dynamics \cite{Olivieri-1995, Travis-1998}, explicit migration cost \cite{Parvinen-2003, Olivieri-1995, Heino-2001}, homogeneous \cite{Travis-1998} or heterogeneous resource distribution \cite{Heino-2001}, static \cite{Travis-1998} or time-varying environments, stochastic environmental dynamics \cite{Olivieri-1995, Heino-2001}, and spatial competition \cite{Cadet-2003} as well as resource competition.  While all of these components are interesting, we focus here on a simpler mechanism for the selection of dispersal.  We claim that in the presence of spatial heterogeneity in resources, no explicit cost of migration or temporal variation in resources is necessary for dispersal to be selected.  Stochasticity, in essence, provides the spatiotemporal heterogeneity to promote dispersal, while spatial variation in resources provides an implicit cost of migration to dissuade from dispersal.  The  combination favors an intermediate dispersal rate.  

Cadet et al \cite{Cadet-2003} examined a continuous time stochastic model with a static, homogeneous environment that incorporated both resource competition, in the form of a density dependent birth rate, and space competition, in the form of maximum population per site which requires newly born individuals to emigrate immediately.  This model incorporates an explicit cost of dispersal by inclusion of dispersers in a pool in which they experience a higher death rate and no possibility of reproduction.  They found that the demographic stochasticity promotes dispersal (distinct from spatial saturation induced emigration) when the patches are, on average, not filled, which allows dispersers an opportunity to improve their lot by leaving dense patches for sparser ones.  They also found a complex dependence of dispersal on the dispersal pool death rate - this is attributed to the fact that, while a higher rate introduces more risk, it also increases the potential reward by reducing the saturation of patches.

Kessler and Sander \cite{Kessler-2009} investigated this phenomenon with a stochastic realization of the Hastings model.  In this model, they found that the faster species could drive the slower one to extinction if the stochastic fluctuations were large (i.e., if there was a small carrying capacity) and/or the spatial variation was small.  They offered a semi-qualtitative estimate, based on the balance between stochastic fluctuations and the implicit cost of dispersal, for the boundary between a fast species victory and a slow species victory. 

In an effort to quantitatively understand this parameter-space boundary between victory of a slower species and that of a faster species, we further examine the relationship between demographic stochasticity and spatial variation in a pair of competing populations.  We introduce, simulate, and analyze a model of two species which differ only in their dispersal rates while competing for spatially-varying resources.  We find that, similar to the spatial model of Kessler and Sander, demographic stochasticity favors the faster disperser, while spatial variation favors the slower.

\section{Approach}
Our intent is to use a simple model which incorporates competition between dispersing species in a heterogeneous environment and which includes intrinsic demographic stochasticity.  In the model, competition exists as resource competition, modeled in the form of density dependent mortality.  Two species will be considered at a time, each with its own dispersal rate.  Though we refer to these rates as ``faster" or ``slower", they do not refer to a movement speed, but rather a propensity for an individual to migrate to a new location.  Spatial variation of resources is modeled as spatially dependent birth rates.  

Such a model must contain two environmental parameters: one which describes the degree of spatial variation in birth rates (or, equivalently, carrying capacities), and one which relates to the spatial average of the carrying capacity.  This latter parameter is an implicit measure of the degree of stochasticity in the model; as the average carrying capacity--and thus the average population--increases, the relative level of stochastic fluctuations decreases.

The model has four parameters: the two dispersal rates and the two environmental parameters.  It thus makes sense to ask two questions regarding the competition.  The first is, for two species with given dispersal rates, how will the result of the competition change as the environmental parameters are varied?  That is, will the faster species drive the slower to extinction first, or vice versa, or will there be coexistence, or \emph {rapid} mutual extinction (extinction being guaranteed in the long term in a stochastic model)?  The second question is, given fixed environmental parameters, is there a \emph{best} dispersal rate?

We use simulations to suggest answers to both questions.  We find that either the faster or the slower species will be preferred, depending upon the parameters --  larger spatial variation and/or larger carrying capacity tends to favor a slower disperser, while smaller values of the parameters tend to favor the faster disperser.  Neither coexistence nor mutual rapid extinction is found.  We also find that, for given environmental parameters, there is an evolutionarily stable dispersal strategy.

Our goal is to find analytical results which explain the findings of the simulations.  This requires several compromises in the details of the model, though the underlying structure is maintained.  Our fundamental approach is to generate systems of differential equations for the moments and correlations of the populations, and then to perform stability analysis on the fixed points of the system.  

However, these moment equations are not closed: lower moments depend on higher moments, which depend on still higher moments, and so on \emph{ad infinitum}.  In order to close some of the moments and make progress, we will consider some simplifying limits of the model.  The most dramatic of these is to assume extremes in the species' dispersal rates: the fast species will disperse infinitely quickly and the slow species  will not disperse at all.  This means that the fast individuals will sample all sites equally and will respond to the mean field.  Importantly, this does not mean that the variance in the fast population goes to zero.  The zero dispersal rate for the slow species allows us to close the correlations between the slow population and the spatially heterogeneous growth rates.

These limits leave us with only the third central moment of the slow population unclosed.  We show that forcefully closing this, by neglecting the third central moment, leads to a quantitatively correct prediction of which species is the victor. It also accurately predicts the trajectories over time of the moments, provided that the slow population does not become too small.

\section{Model}
We consider two species that jointly reside and compete for resources at  spatially distinct sites.  One species is the relatively mobile ``fast" disperser, and the other is the relatively sedentary ``slow" disperser.  The space is inhomogeneous
 in the sense that the carrying capacities of the sites vary.

The following transitions characterize the model process on $L$ sites $i$, where $1 \leq i \leq L$.  $F_i$ is the number of fast-dispersers at site $i$, $S_i$ is the number of slow dispersers, $\gamma_i$ is the growth rate at site $i$.   Individuals experience density dependent mortality scaled by the parameter $\beta = 1/n$; this sets the population scale (which is finite), so that the carrying capacity at a site $i$ is $\gamma_i n$:

\begin{eqnarray}
F_i \xrightarrow{\gamma_i} F_i + 1 & \mbox{  Birth for fast species at site }i
\label{eqn:processstart}\\
S_i \xrightarrow{\gamma_i} S_i + 1 & \mbox{  Birth for slow species at site }i \nonumber\\
F_i \xrightarrow{\beta (F_i+ S_i)} F_i - 1& \mbox{  Logistic death for fast species at site }i \nonumber\\
S_i \xrightarrow{\beta (F_i+ S_i)} S_i - 1& \mbox{  Logistic death for slow species at site }i \nonumber\\
(F_i, F_j) \xrightarrow{D_f / (L-1)} (F_i-1, F_j + 1) & \mbox{  Dispersal for fast species from site }i \mbox{ to site }j\nonumber \\
(S_i, S_j) \xrightarrow{D_s/ (L-1)} (S_i-1, S_j + 1) & \mbox{  Dispersal for slow species from site }i \mbox{ to site }j\nonumber
\end{eqnarray}

For example, the site-dependent growth rate $\gamma_i$ can chosen to be the ``all or nothing" scenario:
\begin{eqnarray}
\gamma_i = 1 & \mbox{ with probability } \phi  &  \mbox{    (fertile sites)}
\label{eqn:gammastart}\\
\gamma_i = 0 & \mbox{ with probability } 1-\phi  &  \mbox{     (sterile sites)}
\label{eqn:processend} \nonumber
\end{eqnarray}
The phenomena studied herein are observed to be insensitive to the specific choice of the $\gamma_i$ provided that there is a tunable spatial variance (which in this case is controlled by the single parameter $\phi$).  See Figure 1 for an illustration of this process.

\begin{figure}
	\centering
		\includegraphics[width=320pts]{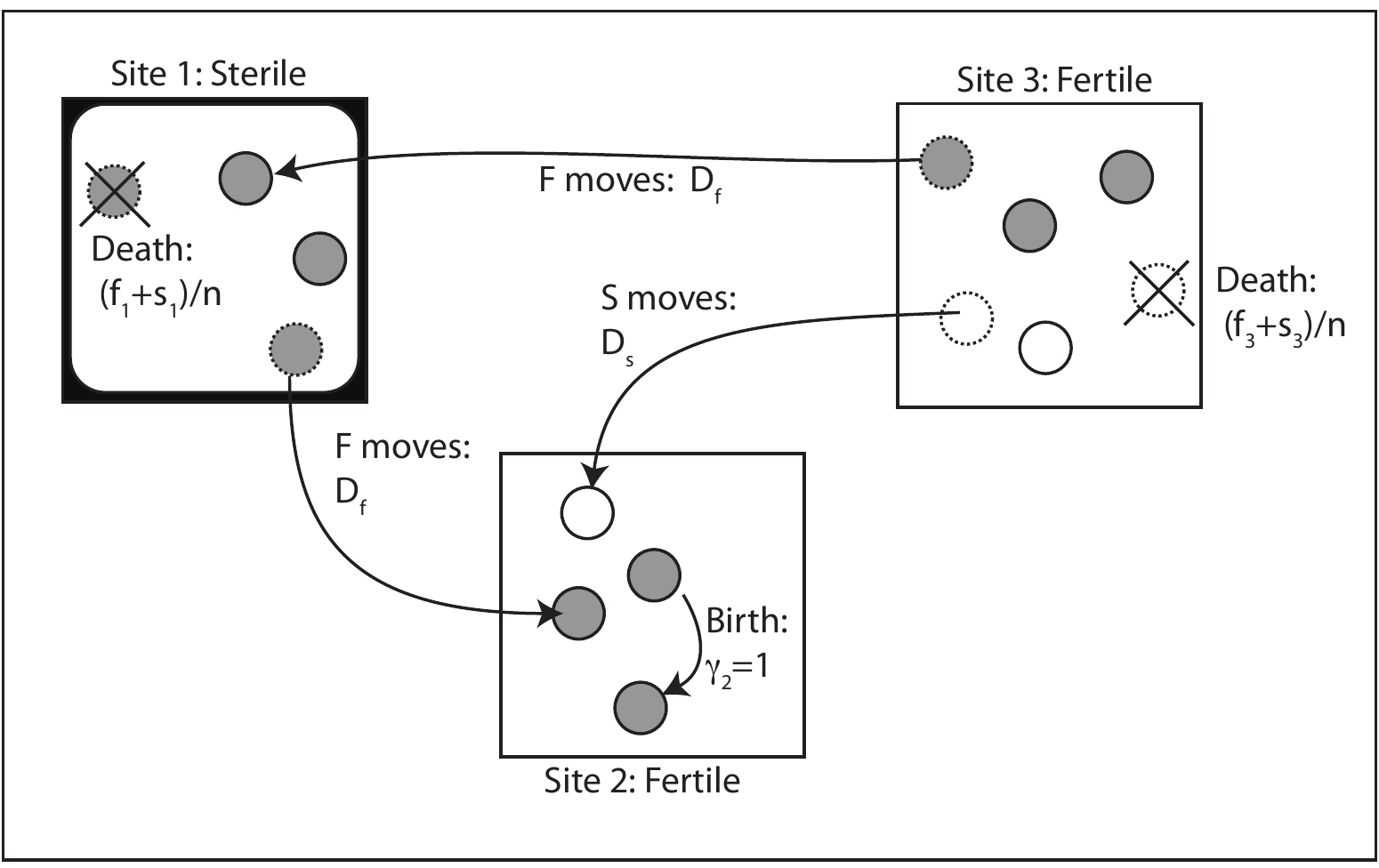}
	\caption{Illustration of the system given by Eqs (\ref{eqn:processstart} - \ref{eqn:gammastart}).  The large squares represent the $L$ sites, and the small (filled, empty) circles represent (fast, slow) individuals.  There are three processes: birth, death, and dispersal.  Birth occurs at rate $\gamma_i = 1$ on fertile sites and $\gamma_i = 0$ on sterile sites (i.e., birth does not occur on sterile sites).  Death occurs at rate $(f_i + s_i)/n$.  Dispersal occurs for $F$ at rate $D_f$ and for $S$ at rate $D_s$.}
	\label{fig:cartoon}
\end{figure}

We studied this process in two ways and compare the results.  First, the process was simulated exactly, producing realizations of the Markov process according to its definition.  We refer to these as the direct Monte Carlo simulations (DMCS).  Secondly, we analyze the differential equations of the moment hierarchy generated by the process.

In order to make analytical progress we examine the simultaneous limits of many sites, infinitely fast dispersal for the fast species, and no dispersal for the slow, i.e. $L \to \infty$, $D_f \to \infty$, and $D_s \to 0$.  It is important to note that we do \emph{not} take the infinite $n$ limit in which the process would be deterministic.  This analysis confirms that the demographic stochasticity due to finite $n$ is important, as demonstrated by DMCS.

\section{Simulations}
\begin{figure}
	\centering
		\includegraphics[width=250pts]{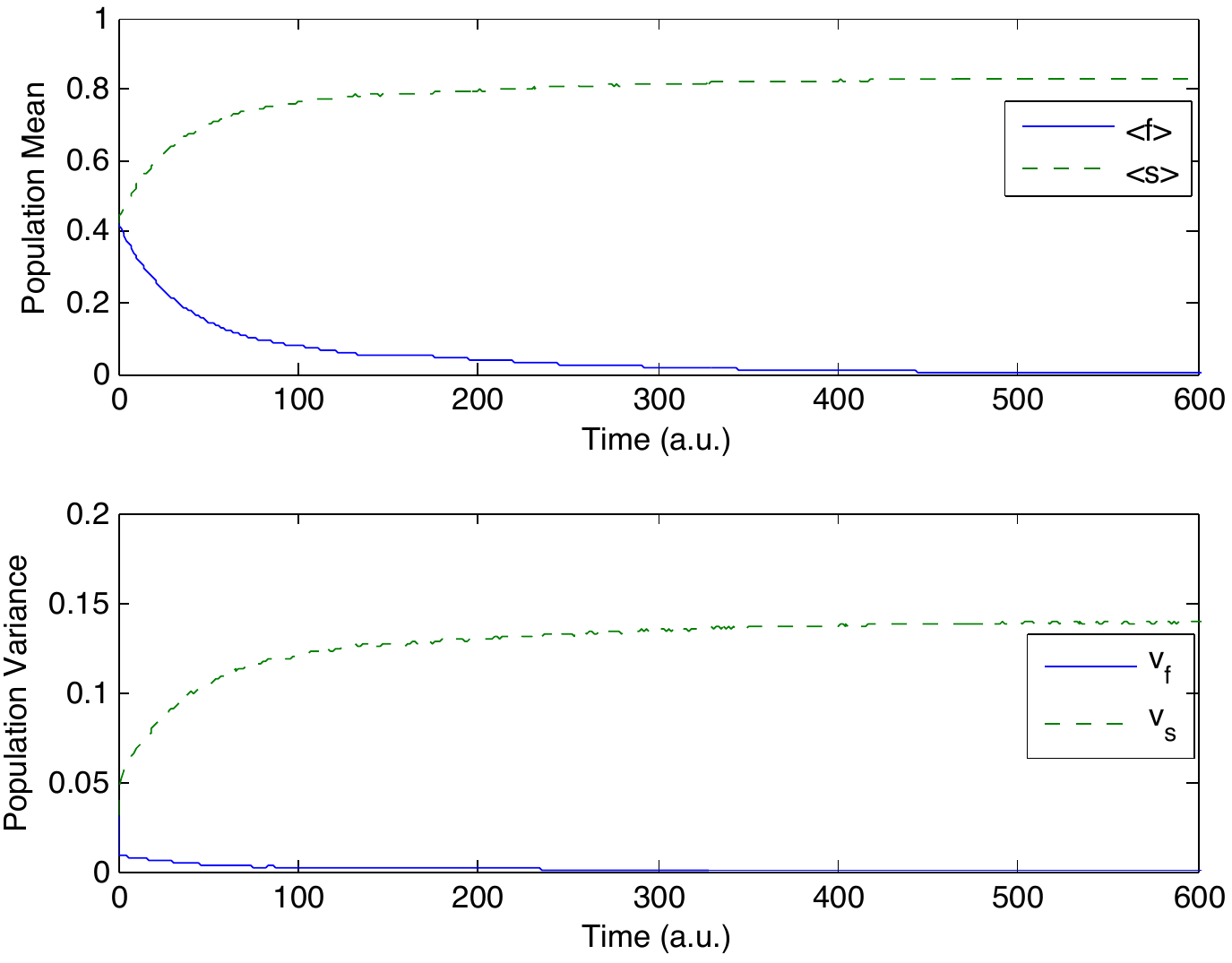}
	\caption{Example mean trajectory from simulations.  $\la f \ra$ and $\la s \ra$ are the mean populations normalized by $n$, and $v_f$ and $v_s$ are the normalized population variances, over $L=100$ sites and 100 simulations.  For these simulations, $n = 50$, $\phi = 0.85$, $D_f = 10$, and $D_s = 0$.  In this scenario, S drives F to extinction.}
	\label{fig:trajectory}
\end{figure}

For each particular simulation, the parameters of number of sites $L$, the fraction of fertile sites $\phi$, the population scale $n$, and the dispersal rates $D_f$ and $D_s$ were fixed.  Initial conditions were that every fertile site started with an equal number of fast and slow individuals, half of the local carrying capacity for each (the carrying capacity took only even values), and that all sterile sites were initially empty.  The time course of the populations on each site were recorded for each, as was the eventual winner, i.e., the species that drove the other to extinction (no coexistence in any realization was observed for $D_f \neq D_s$).  By averaging over many simulations,  trajectories were generated for the moments of the populations (cf. Fig \ref{fig:trajectory}).  The fraction of victories for each species was also recorded for each given set of parameters.

\begin{figure}
	\centering
		\includegraphics[width=250pts]{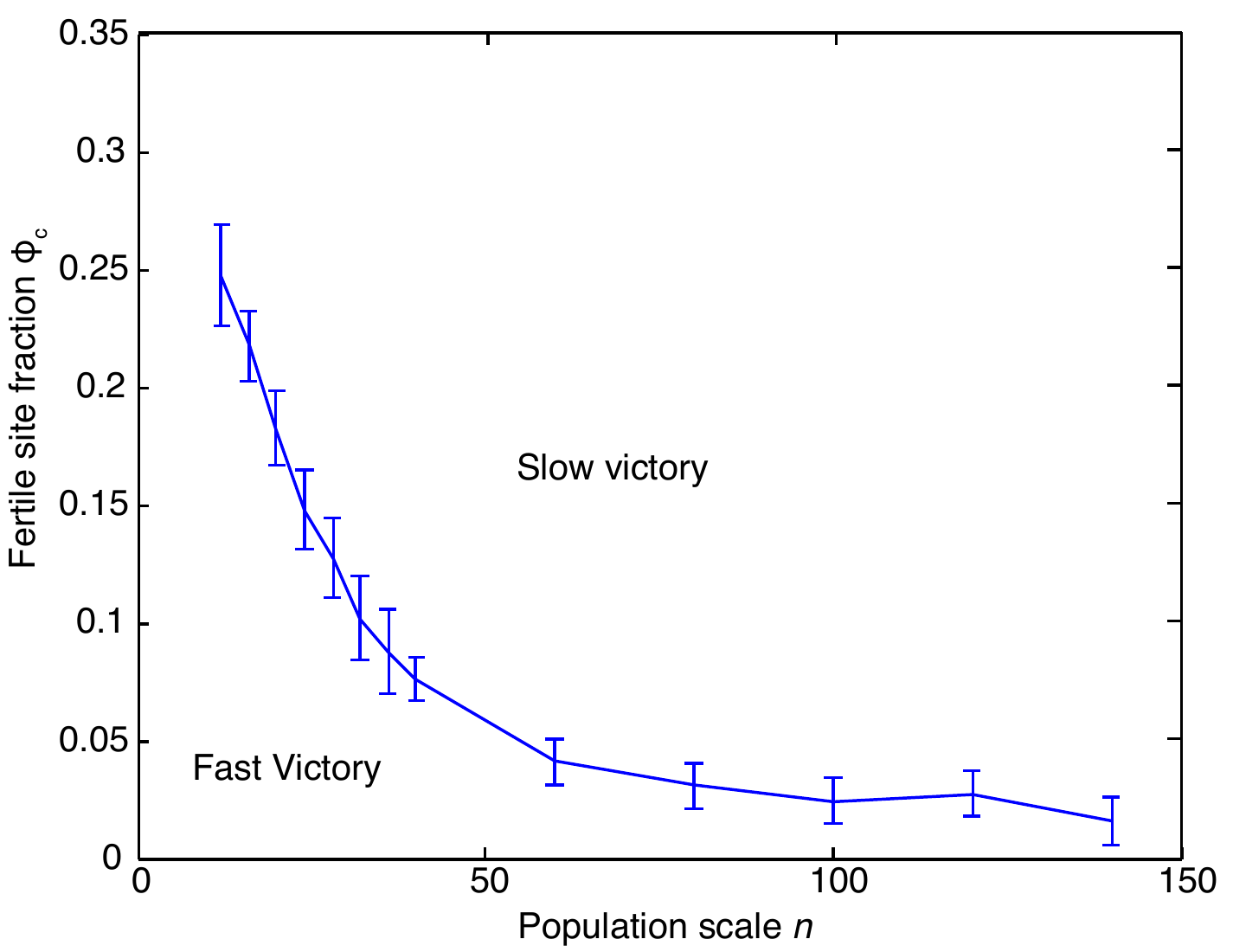}
	\caption{Phase boundary between victory by the slow species (above) and victory by the fast species (below).  The points are the result of continous-time rejection-free Monte Carlo simulations, with the dispersal rates $D_f = 10$ and $D_s = 0$.}
	\label{fig:boundaryInitial}
\end{figure}

While varying only the fraction of fertile sites $\phi$ and the population scale $n$, two distinct phases emerge, separated by a boundary (Fig \ref{fig:boundaryInitial}).  Below this boundary, the faster species wins, while above the slower species wins.  This is in contrast with the results for the deterministic $n \to \infty$ limit, where the slower species always wins.  While DMCS recovers the deterministic result for large $n$, the faster species wins when the population scale $n$ or spatial variance of growth rates $Var(\gamma) = \phi(1-\phi)$ is small enough.  This phenomenon was first explicitly noted by Kessler and Sander \cite{Kessler-2009}.

Kessler and Sander made a semi-quantitive argument for the form of the fast/slow phase boundary based on the competition between the degree to which each population fluctuates and the degree to which each population best follows the contours of the food distribution.  This argument determined that the phase boundary should scale as the following: 
\begin{equation}
Var(\gamma) \sim 1/n
\label{eqn:Boundary}
\end{equation}
where $n$ is the average carrying capacity.  The goal of the following section is to develop a quantitative understanding for this phase boundary.


\section{Analysis}
From the Master equation corresponding to (\ref{eqn:processstart}) and (\ref{eqn:gammastart}) we generate differential equations for the evolution of the moments.  Let $f_i$ be $F_i/n$.  In order to make the averaging precise, we consider the mean-field limit where $L$ goes to infinity.  We define the averages and the variances according to:

\begin{equation}
\la f \ra = \mathbb{E}(\lim_{L \to \infty} \frac{1}{L} \sum_{i=1}^L f_i)
\end{equation}
\begin{equation}
v_f = \la f^2 \ra - \la f \ra^2 = \mathbb{E} \big (\lim_{L \to \infty} \frac{1}{L} \sum_{i=1}^L (f_i - \la f \ra)^2 \big)
\end{equation}
Similar definitions hold for $S$ and high moments of the random variables, leading to the following moment equations:

\begin{eqnarray}
\frac{d \la f \ra}{dt} &=& \la \gamma f \ra - \la f s \ra - \la f \ra^2 - v_f \\
\frac{d v_f }{dt} &=& 2 \big( \la \gamma f^2 \ra - \la \gamma f \ra \la f \ra + \la f \ra^3 + \la f \ra v_f - \la f^3 \ra + \la f \ra \la f s \ra - \la f^2 s\ra \big) \nonumber \\  &&- 2 D_f (v_f  - \frac{\la f \ra}{n}) + \big(\la \gamma f \ra + \la f \ra ^2 + v_f + \la f s \ra \big)/n
\end{eqnarray}

and 

\begin{eqnarray}
\frac{d \la s \ra}{dt} &=& \la \gamma s \ra - \la f s \ra - \la s \ra^2 - v_s \\
\frac{d v_s }{dt} &= &2 \big( \la \gamma s^2 \ra - \la \gamma s \ra \la s \ra + \la s \ra^3 + \la s \ra v_s - \la s^3 \ra + \la s \ra \la f s \ra - \la s^2 f\ra \big) \nonumber \\ &&- 2 D_s (v_s - \frac{\la s \ra}{n}) + \big(\la \gamma s \ra + \la s \ra ^2 + v_s+ \la f s \ra \big)/n,
\label{eqn-SEq}
\end{eqnarray}

and the correlation equation:
\begin{equation} 
\frac{d \la fs \ra}{dt} = 2 \la \gamma f s \ra - 2 (\la s f^2 \ra + \la f s^2 \ra) +(D_s + D_f) (\la f \ra \la s \ra - \la fs \ra ).
\end{equation}

Now we focus on the limits of $D_f\to\infty$ and $D_s \to 0$.  In the large $D_f$ limit, there is rapid relaxation to $v_f = \la f \ra/n$.  This frees us from the need of a closure for the $\la f^3 \ra$ and $\la f^2 s \ra$ terms.   Similarly, $\la fs \ra$ rapidly relaxes to $\la fs \ra = \la f \ra \la s \ra$.  The simulations confirm the accuracy of these approximations; even for $D_f = 10$, the error was only a few percent.

The equations for the higher correlations are readily computed, but lengthy.  Note in particular the following:
\begin{eqnarray}
\frac{d \la s^2 f \ra}{dt} =D_f \big(\la f \ra \la s^2 \ra - \la s^2 f \ra \big) + O(1).
\end{eqnarray}
In the large $D_f$ limit, there is also a rapid relaxation of $\la s^2 f \ra$ to $\la f \ra \la s^2 \ra$.  This closure approximation was also verified by the simulations, and is found to be accurate even for moderate $D_f$.  Then taking $D_s \to 0$ we eliminate the $D_s$ term in (\ref{eqn-SEq}).

Next, the correlations $\la \gamma s^m \ra$ for integer $m$ are calculated using the ``all or nothing" definition of gamma (\ref{eqn:gammastart}).
Since $D_s = 0$, a site with no slow individuals remains devoid of them for all future times.  Therefore, given initial conditions where populations start only on the fertile sites,
\begin{equation} 
\la \gamma s \ra = 1/L \sum_{i=1}^{L} s_i = 1/L \sum_{i = 1}^{L = \phi L} s_i + 1/L \sum_{i=\phi L + 1}^{L} 0 = \la s \ra.
\end{equation}
This applies similarly for all moments involving $\gamma$ and $s$.  That is, $\la \gamma s^2 \ra = \la s^2 \ra$, and so on.
However, $F$ moves about rapidly across all the sites, sampling them equally in the large $D_f$ limit.  Thus the correlation $\la \gamma f\ra = \phi \la f\ra$, which the simulations confirm as well at even moderately large $D_f$.

We can then simplify the equations to the form
\begin{eqnarray}
\frac{d \la f \ra}{dt} &=&  \big( \phi  - \la f \ra - \la s \ra - 1/n \big)\la f \ra \\
\frac{d \la s \ra}{dt} &=& (1 - \la f \ra - \la s \ra - v_s / \la s \ra \big) \la s \ra\\
\frac{d v_s }{dt} &=& 2 \big(v_s+ \la s \ra^3 + \la s \ra v_s - \la s^3 \ra)  + \big(\la s \ra + \la s \ra ^2 + v_s+ \la f \ra \la s \ra \big)/n
\end{eqnarray}

It is now convenient to make a change of variables.  Let $p(t)$ be the mean of the slow population on the fertile sites.  Since the slow population is zero on the sterile sites, $\la s \ra = \phi p$.  The variance $v_s$ is comprised of two components - the variance on the fertile sites $v_p$, and the variance between the populated fertile sites and the abandoned sterile sites.  Thus, $v_s = \phi v_p + p ^2 \phi (1- \phi)$.  For convenience, we apply the same scaling with the fast population so that $\la f \ra = \phi q $ defines the new variable $q(t)$.  

Let $\xi$ be the 3rd central moment of the slow population on the fertile sites.  That is, $\xi = \la (p_i - p)^3 \ra$, where $p_i$ is the random value of the slow species population on each fertile site, so that $p$, defined above, can be written as $p = \la p_i \ra$. 

The equations then become
\begin{eqnarray}
\frac{d q}{dt} &= & \phi \big( 1  - q - p - 1/(n \phi) \big)q
\label{eqn:finala}\\
\frac{d p}{dt} &= & \big(1 - \phi q - p - v_p/p \big)p\\
\frac{d v_p}{dt} &= &2 \big(1 - \phi q -2 p + 1/(2 n) \big)v_p  +  (1 + p  + \phi q ) p/n  - 2 \xi
\label{eqn:final}
\end{eqnarray}

\begin{figure}
	\centering
		\includegraphics[width=250pts]{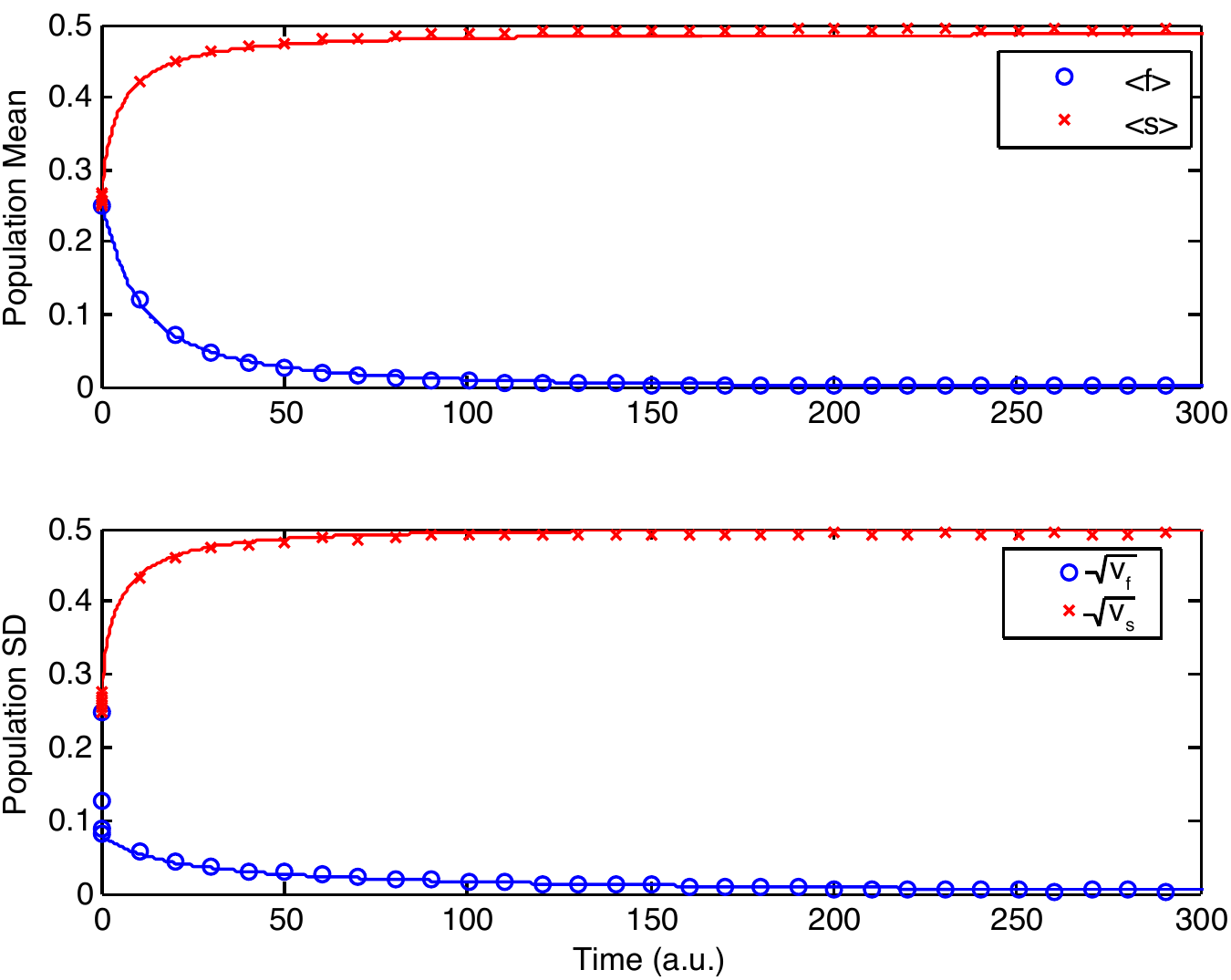}
	\caption{Example of the mean populations from the ODEs and the simulations when S drives F to extinction.  $\la f \ra$ and $\la s \ra$ are the mean populations, and $v_f$ and $v_s$ are the population variances.  The solid lines are given by solving the ODEs (or, in the case of $v_f$, by the closure used in the ODEs), and the points (circles for $F$ and crosses for $S$) represent the moments over 100 simulations.  For these simulations, $n = 40$, $L=100$, $\phi = 0.5$, $D_f = 10$, and $D_s = 0.001$.}
	\label{fig:trajectoryODE}
\end{figure}

\begin{figure}
	\centering
		\includegraphics[width=250pts]{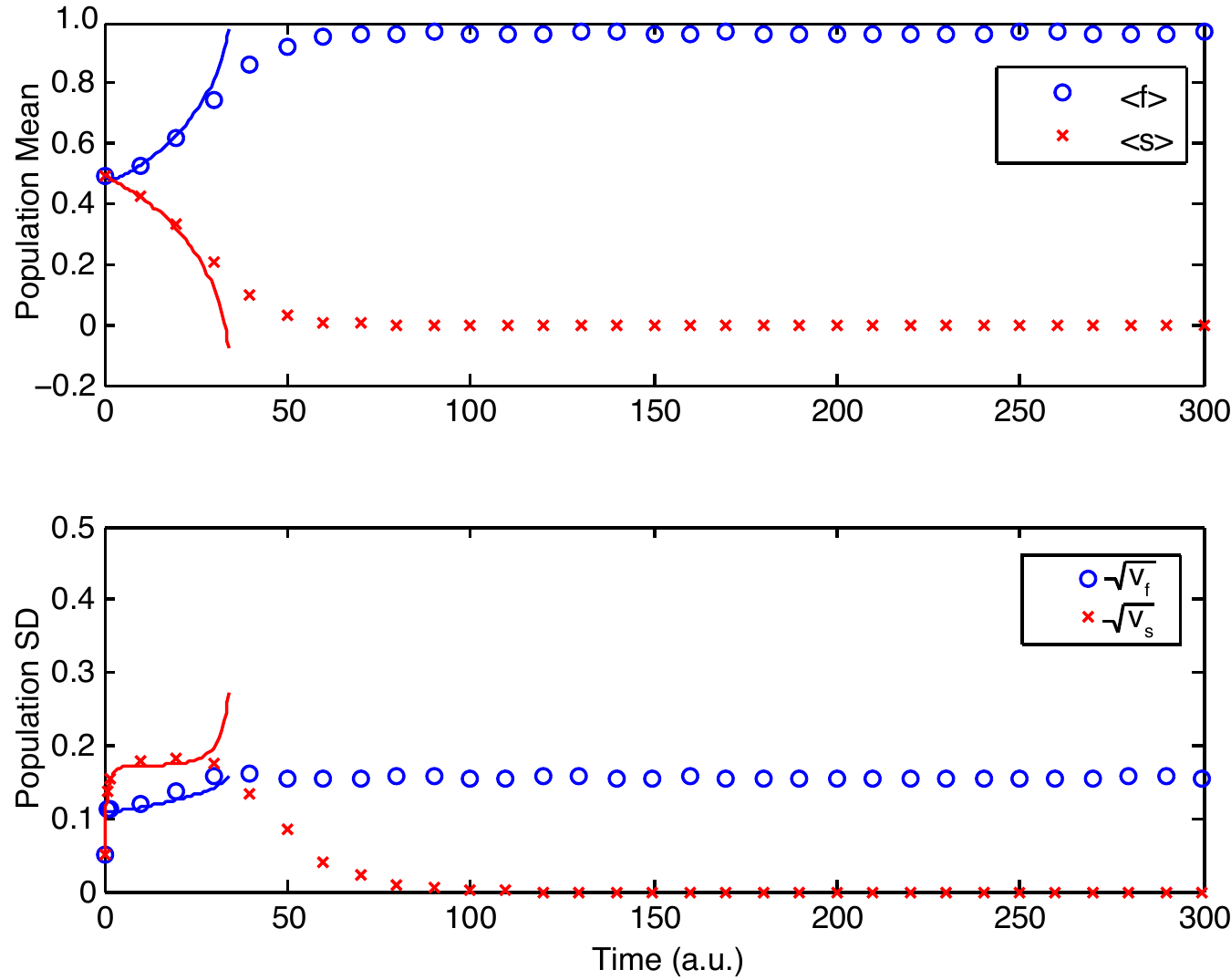}
	\caption{Example of the mean populations from the ODEs and the simulations when F drives S to extinction   Note that the ODE trajectory initially matches the simulations until suddenly it diverges and S becomes negative, at which times the ODE integration is ceased.  Although the trajectories diverge at this point, the ODEs have made a prediction for the outcome: S will go extinct.  For these simulations, $n = 40$, $L=100$, $\phi = 0.99$, $D_f = 10$, and $D_s = 0.001$.}
	\label{fig:trajectoryODE2}
\end{figure}

The remaining challenge is the third moment $\xi$, which is not closed.  In circumstances when $p$ is large, such as when the slow population is dominating, we expect that $\xi$ is small compared to the other terms and can be safely dropped.  The simulations bear out this expectation.  Dropping the last term in (\ref{eqn:final}) leads to an accurate prediction when the slow population wins (see Figure \ref{fig:trajectoryODE}).
When the slow population is near extinction, $\xi$ is not small compared to the other terms in the $v_p$ equation and it cannot be safely neglected.  Neglecting $\xi$ in these situations leads to instability of the ODE system: the evolution drives $v_p$ to unphysical negative values and $p$ diverges (see Figure \ref{fig:trajectoryODE2}).

Thus, while neglecting $\xi$ does not always reproduce the results of the stochastic system, it still leads to an accurate prediction of the final outcome.  When the ODEs predict that the slow population will win, then the ODEs are well behaved and produce good quantitative approximations.  When the ODEs demonstrate instability, then $\xi$ is not small, indicating that the slow population is going extinct, i.e. that the fast population is winning.  That is to say, the model always offers a quantitative prediction of the binary variable describing whether $F$ or $S$ is the ultimate victor, even if it does not always quantitatively predict the moments as a function of time.

When $\xi$ is neglected to close the system, Eqs. (\ref{eqn:finala} - \ref{eqn:final}) have four pertinent fixed points.  A finite stochastic system like this will eventually end in total extinction, the only absorbing state, at a time that is exponentially large in the population.  Rate equations are not sensitive to these large deviations, and stochastic differential equations get quantitatively wrong results, so more sophisticated techniques would have to be used to determine the extinction time \cite{Doering-2005}.  This ultimate extinction occurs in the simulations, which of course require finite $L$, but because $L$ is large the extinction time is very long compared to the time scale of a single-species victory.  The ODE analysis is for the case that $L \to \infty$, in which case the extinction time diverges, so this aspect is not contained in the ODE model.
 
The four fixed points are:
\begin{eqnarray}
(q, p, v_p) = & (0, 0, 0) &\mbox{Extinction}\\
= & (1 - \frac{1}{n\phi}, 0, 0) &\mbox{Victory by fast}\\
= & (0, \frac{1}{4}(3+ \sqrt{\frac{n-8}{n}}), \frac{1}{8}(1-\sqrt{\frac{n-8}{n}}+\frac{4}{n})) &\mbox{Victory by slow}\\
= &  (0, \frac{1}{4}(3- \sqrt{\frac{n-8}{n}}), \frac{1}{8}(1+\sqrt{\frac{n-8}{n}}+\frac{4}{n})) &\mbox{Saddle}
\end{eqnarray}
These expressions are derived for large $n$, and only make sense for $n > 8$.  Neither the extinction nor fast-victory fixed points are ever stable since both have two strictly positive eigenvalues.

\begin{figure}
	\centering
		\includegraphics[width=250pts]{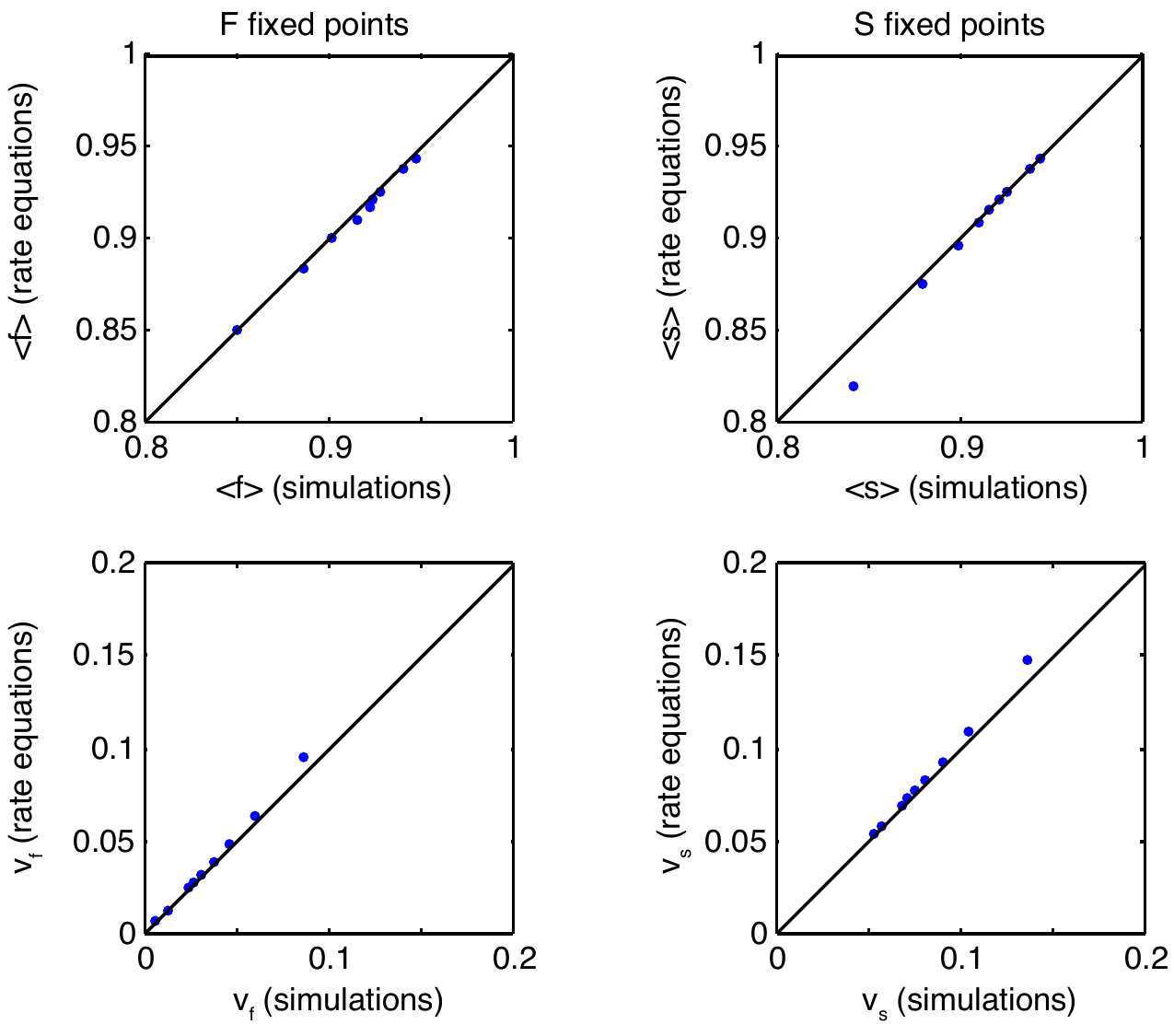}
	\caption{Fixed points of the rate equations versus the quasi-steady state moments from the simulations.  In each case, the competing species population was set to zero, so the equation follow the competition-free trajectory to the quasi-steady state.  The mean and variance of both the fast and slow species' populations were well matched by the rate equations, particularly as the population scale $n$ increased.  These were computed with $D_f=10$, $D_s=0$, $L=100$ and $\phi = 0.95$, for $n$ ranging from $10$ to $160$.  }
	\label{fig:fxpts}
\end{figure}

The fast species achieves victory with some parameters in the stochastic system, but because this requires a small slow population and $\xi$ can no longer be safely neglected, the rate equations do not ``find" this result.  Despite the inaccuracy in the prediction of the stability, the fixed points do give the quasi-steady state value achieved in the simulations (to within an error of $1/L$.  See Figure \ref{fig:fxpts}).

The slow-victory fixed point has two eigenvalues that are always negative.  The remaining (largest) eigenvalue is also negative when 
\begin{equation}
n > \frac{2}{\phi(1-\phi)} = \frac{2}{Var(\gamma)},
\label{eqn:sstab}
\end{equation}
equivalent to the expression determined in the Kessler and Sander model.  Again, the value of the stochastic quasi-steady state is achieved by the equations.

\begin{figure}
	\centering
		\includegraphics[width=250pts]{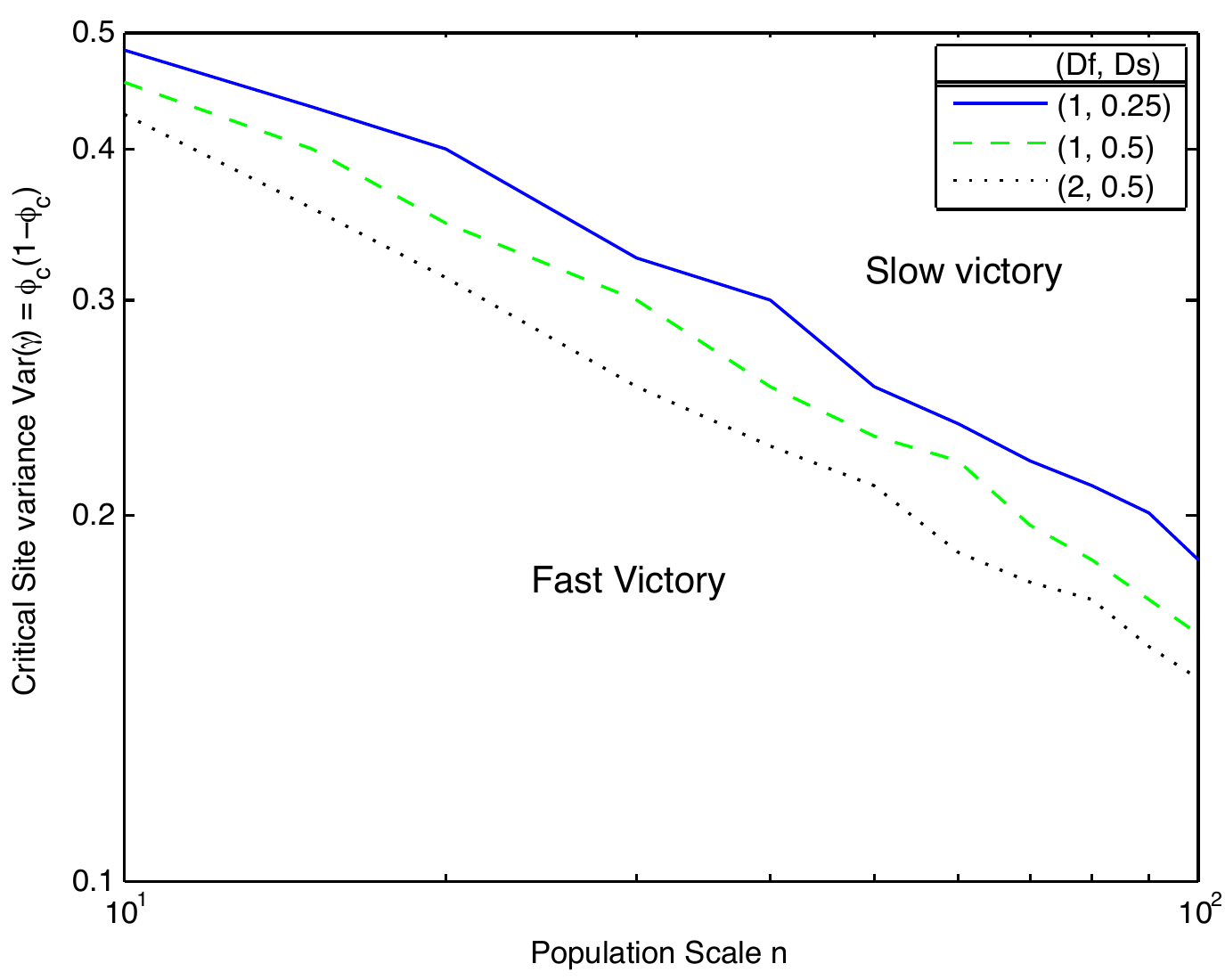}
	\caption{Phase boundary between victory by the slow species (above) and victory by the fast species (below).  The points are the result of continous-time rejection-free Monte Carlo simulations with dispersal rates $D_f = 10$ and $D_s = 0$.  For each set of parameters, 100 simulations were conducted.  The points indicate the parameters which generate a 50/50 split for victories between S and F, and the error bars represent a 95\% confidence interval.  The solid line is the result of integrating the differential Eq. (\ref{eqn:final}), and then decrementing $n$ for each value of $\phi$ .  The dotted line marks the stability threshold of the s-victory fixed point (\ref{eqn:sstab}) of the differential equations, which is a lower bound for the phase boundary.  The differential equations match the simulations well, particularly in the high carrying capacity limit.}
	\label{fig:boundary}
\end{figure}

The final fixed point, with at least one positive and one negative eigenvalue, serves to split the phase space into the domain of attraction for the slow-victory fixed point.  It is important to note that when the slow-victory fixed point is stable, it is the initial conditions combined with the extent of the basin of attraction that determine whether a specific trajectory will result in a slow or fast victory.  That is, $n > \frac{2}{\phi (1-\phi)}$ is a necessary, but not sufficient, condition for the slow species victory.  Thus the stability threshold $n =\frac{2}{\phi (1-\phi)}$ gives a lower bound to the phase boundary (see Figure \ref{fig:boundary}).

It is noteworthy that the stability condition (\ref{eqn:sstab}) yields a relationship similar to that estimated by Kessler and Sander for the boundary.  The estimate goes as follows: by taking the difference in the rate equations for the means, we find:

\begin{equation} 
\frac{d \la f - s \ra}{dt} = \la f - s \ra \big[\la \gamma \ra - \la f + s \ra \big] + \big[ \mbox{Cov}(\gamma, f) - \mbox{Cov}(\gamma, s)\big] - \big[v_f - v_s \big]
\end{equation}
where Cov$(\gamma, f) = \la \gamma f \ra - \la \gamma \ra \la f \ra$.
If the two populations are the same, the sign of the change depends on the competition between two differences.  The fast population smoothes its profile, reducing both its variance and its covariance with the $\gamma_i$. Approximating these as zero leaves the difference between $v_s$ and covariance of $s$ and $\gamma$ to determine the advantage.   The population $s$ is of order $ \gamma_i$, so Cov$(\gamma, s) \approx   \mbox{Var}(\gamma)$.  When $\phi$ is near unity, the variance $v_s \approx \phi/n$ due to demographic stochasticity.  Equating Cov$(\gamma, s)$ and $v_s$ implies Var$(\gamma) = \phi/n$, which in turn implies:
\begin{eqnarray}
n = \phi/\mbox{Var}(\gamma)
\end{eqnarray}
which is similar to the stability condition given by (\ref{eqn:sstab}) when $\phi$ is near unity.

\section{Summary and Discussion}
We have developed a model to investigate the effect of intrinsic demographic stochasticity on competition between dispersers in a heterogeneous environment.  In order to get analytic results, we examined extreme limits in dispersal rates, which allowed the closure of most of moment equations, leaving only the third central moment of $S$ unclosed.  Simply neglecting this moment led to accurate quantitative predictions of the population trajectories over time when $S$ remained large, and regardless of the size of $S$, to an accurate quantitative prediction of the outcome, i.e. which species drives the other to extinction.

Via simulation and analysis we have confirmed the observation that \emph{intrinsic demographic stochasticity} can favor the fast dispersal in head-to-head competition with a slower competitor.  This can be intuitively interpreted in the following way: demographic stochasticity leads to sites which are, by chance, inhabited by fewer individuals than the carrying capacity allows.  A fast disperser is more likely than a slow disperser to happen upon these sites, after which it can take advantage of the remaining capacity \cite{Parvinen-2003}.


This is to be contrasted with the deterministic results of Hastings and of Dockery et al. that spatial variations in fecundity drive dispersal rates lower \cite{Hastings-1983, Dockery-1998} due to an implicit cost of dispersing for the faster disperser when it leaves favorable habitats.  In these deterministic works, dispersing individuals have no opportunity happen upon sites that are by chance favorable.  The cost of dispersal without benefit drives dispersal rates to zero.

Explicit dispersal costs also drive down dispersal rates \cite{Hamilton-1977, Comins-1980}.  As dispersal benefits are removed, for example by reducing kin competition or demographic stochasticity by increasing the carrying capacity, slower dispersal is again favored.    If dispersal benefits are held constant as dispersal costs are reduced, either by lowering the explicit dispersal cost or by reducing the spatial variation in the environment, higher dispersal rates are favored.

Other factors are known to influence dispersal rates.  Kin competition \cite{Hamilton-1977}, habitat extirpation risk \cite{Comins-1980} and temporal habitat variability \cite{Johnson-1990, Dieckmann-1999, Hutson-2001, Baskett-2007} favor faster dispersal.

In the absence of factors detrimental to dispersal, dispersal rates in a stochastic model increase.  In our model, if the environment is spatially homogeneous, dispersal is not a neutral factor, as it was in Hastings model.  There is no cost to dispersal, implicit or explicit, but there is a benefit to dispersal.  This leads to a faster disperser always defeating a slower disperser (data not shown; for finite $L$, the two species reach parity asymptotically if both dispersal rates are much faster than the birth rate, though cohabitation is not observed).

\begin{figure}
	\centering
		\includegraphics[width=250pts]{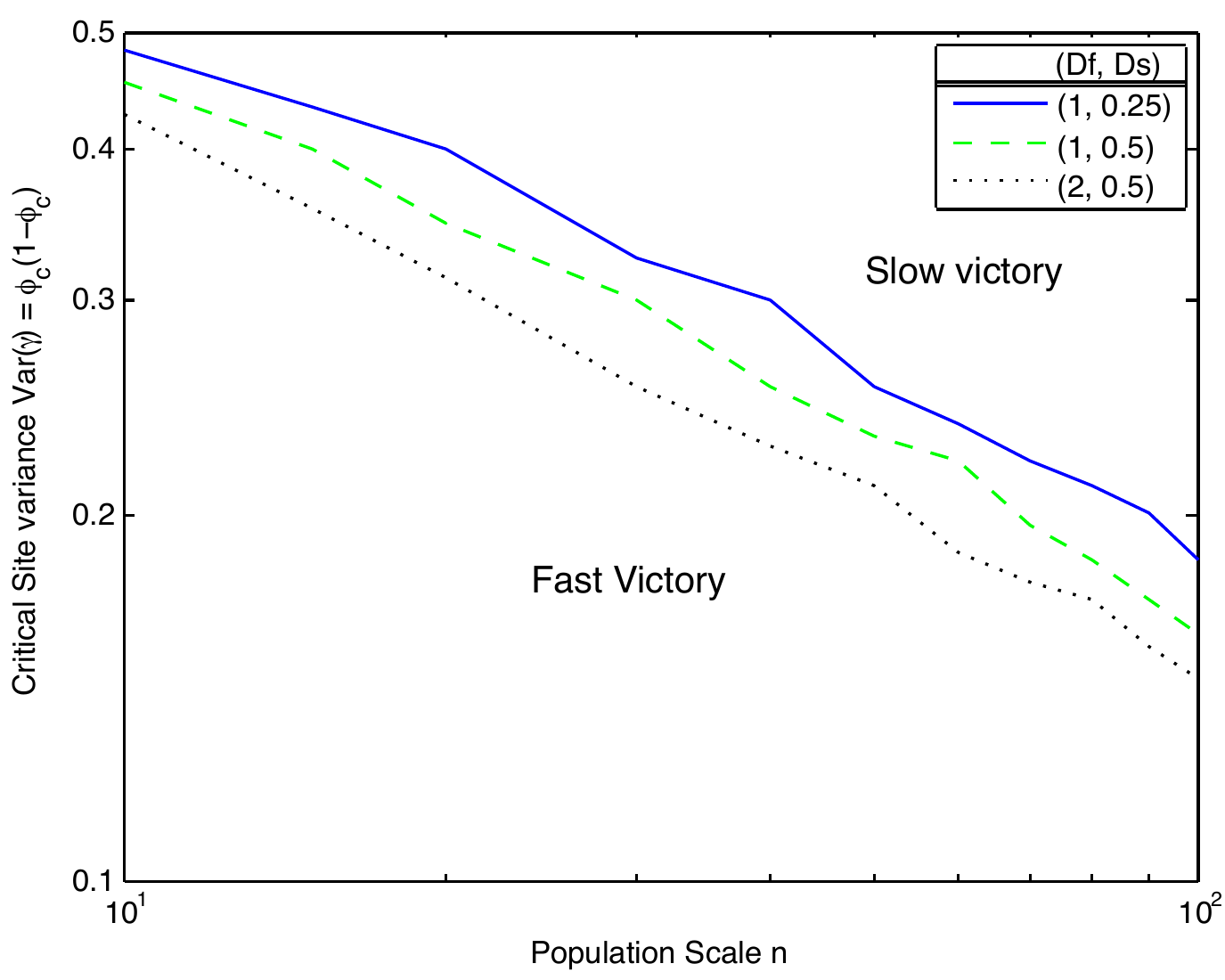}
	\caption{Phase boundary between victory by the slow species (above) and victory by the fast species (below) for three pairs of values for dispersal rates $D_f$ and $D_s$.  The qualitative forms of the boundaries are similar, and also similar to figure \ref{fig:boundaryInitial}.  However, the boundary shifts as the dispersal rates $D_f$ and $D_s$ are varied.  This is to be expected if there exists an optimal dispersal rate which depends on the population scale and spatial variation in habitat quality.  For example, if one species has a dispersal rate that is fixed at the optimum for a given population scale and spatial variation, then that point on the phase diagram should be in the slow-victory phase if the competing species has a faster dispersal rate, but in the fast-victory phase if the competing species has a slower dispersal rate.}
	\label{fig:boundaryModDs}
\end{figure}

\begin{figure}
	\centering
		\includegraphics[width=250pts]{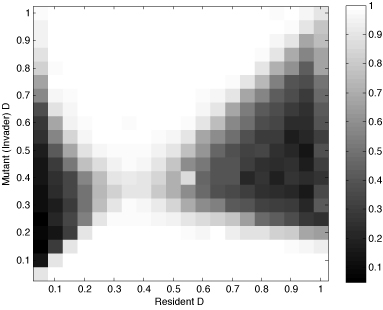}
	\caption{Pairwise invasibility plot of dispersal.  The resident species begins at carrying capacity on each fertile site, and the invading species begins with one individual on each fertile site.  Color represents the fraction of trials in which the invading species goes extinct.  An optimum dispersal of $D^*\approx0.4$ is evident.  In this example, 100 simulations were run for each point, and $n = 10$, $L = 500$, and $\phi = 0.5$.}
	\label{fig:Optimum}
\end{figure}

When dispersal-promoting and dispersal-suppressing factors compete it is expected that an optimal dispersal rate might exist, the rate that best balances the factors and thus defeats any competitor with either a higher or a lower dispersal rate.  There is numerical evidence in these simulations (away from the extreme limits in the analysis section) that this is the case in the model introduced here as well.   The boundary, which has the same qualitative form for dispersal rates less extreme than the limits taken in the analysis section (Figure \ref{fig:boundaryModDs}), implies that the optimum dispersal rate decreases when the population scale $n$ or the variance of the site quality Var($\gamma$) increases, since the slower species wins in these limits.  We see this explicitly by varying both dispersal rates, and discovering that, for the parameters used, the species with $D \approx 0.35 $ can both invade a resident population and resist invasion by a mutant (Figure \ref{fig:Optimum}).

While it is challenging to make quantitative comparisons between the parameters in our model and empirical observations, in some cases the predictions seem to agree with real data.  For example, Lowe  \cite{Lowe-2009} compared the long distance dispersal rates of salamanders measured by kurtosis of movement distribution with the goodness of habitat measured by body condition of the salamanders.  It was found that dispersal decreased when the average body condition was high (analogous to large $n$) or when body condition variance was high (analogous to large Var$(\gamma)$), precisely as expected.


It is also of interest to note that the system studied in this paper is one for which the effects of stochastic fluctuations have a qualitative impact on the result for \emph{arbitrarily large} population scale $n$.  While it is true for fixed spatial variation Var$(\gamma)$, $n$ may be chosen to be large enough to recover the deterministic result (viz., that the slow species wins), there is no value of $n$ which is large enough to always ensure a deterministic result for arbitrary Var$(\gamma)$.  That is, for any $n$, there exists a Var$(\gamma) < 2/n$ small enough such that the effects of demographic stochasticity leads to a fast species victory.  

\section{Acknowledgments}
This project was supported in part by the National Science Foundation Award DMS 0554587. The Center for the Study of Complex systems at the University of Michigan provided computing time.


\section{References}
\bibliographystyle{elsarticle-num.bst}

\end{document}